\documentclass{article}
\usepackage{graphicx}
\usepackage{amsfonts}
\usepackage{amsmath}
\usepackage{amssymb}
\usepackage{bm}
\usepackage{wrapfig}
\usepackage{txfonts} 

\newtheorem{theorem}{Theorem}
\newtheorem{remark}{Remark}
\newcommand{\ua}{U^\alpha}
\newcommand{\ub}{U^\beta}

\newcommand{\inta}{\int_{\R^3} \int_0^\infty }
 \newcommand{\II}{\mathcal{I}}
 \newcommand{\R}{\varmathbb{R}}

\newcommand{\ug}{U^\gamma}
\newcommand{\la}{\lambda^\alpha}
\newcommand{\llg}{\lambda^\gamma}
\newcommand{\lb}{\lambda^\beta}
\newcommand{\qg}{q^\gamma}
\newcommand{\qb}{q^\beta}
\newcommand{\tbg}{t^{<\beta \gamma>_3}}
\newcommand{\gab}{g^{\alpha\beta}}
\newcommand{\gag}{g^{\alpha\gamma}}
\newcommand{\gbg}{g^{\beta\gamma}}

\newcommand{\hab}{h^{\alpha\beta}}
\newcommand{\hag}{h^{\alpha\gamma}}
\newcommand{\hbg }{h^{\beta\gamma}}

\newcommand{\Ua}{U_\alpha}
\newcommand{\Ub}{U_\beta}
\newcommand{\Ug}{U_\gamma}

\newcommand{\Hbg}{h_{\beta\gamma}}

\newcommand{\Hbm}{h_{\beta\mu}}

\newcommand{\Hbd}{h_{\beta\delta}}

\newcommand{\Hgn}{h_{\gamma\nu}}

 \newcommand{\Hmn}{h_{\mu\nu}}

\begin{document}

\title{Eckart  equations, Maxwellian  iteration  and    Relativistic  Causal  Theories  of Divergence type}

\author{
Tommaso Ruggeri\\
Mathematical Department, University of Bologna and Accademia dei Lincei\\
e-mail: tommaso.ruggeri@unibo.it} 
\date{}
\maketitle





\begin{abstract}
	We consider a general causal   relativistic theory of divergence type in the framework of  Rational Extended Thermodynamics (RET)
 for a  compressible, possibly dense, gas. We require that the system converges in the Maxwellian iteration's first step to the parabolic Eckart equations. This requirement implies a constraint between the two coefficients present in the triple tensor evaluated at equilibrium. Moreover, the production tensor is determined for prescript thermal and caloric state equations and given heat conductivity, shear, and bulk viscosities. In the second part, we prove that if the original hyperbolic system satisfies the universal principles of RET,  as can be put in the symmetric form using the \emph{main field}, it   always satisfies the previous compatibility condition. Therefore any causal   system  of divergence type that satisfies the entropy principle with a convex entropy converges to the Eckart system in the Maxwellian iteration also when we have no information at the mesoscopic scale from the kinetic theory.
 The obtained results are tested on the RET theories of rarefied monatomic and polyatomic gases.
 \end{abstract}



\section{Introduction}
As it is well known, the pioneering papers by M\"uller \cite{Mull} and Israel \cite{Isr} are the first tentative to obtain a causal   relativistic phenomenological theory with a system of equations of   hyperbolic type such that the  wave speeds are finite consistently with  the relativity principle. This approach is based substantially on the idea of the use of the modified Gibbs relation in a non-equilibrium state. This method has been fundamental for a long time because of its simplicity, but more refined analyses reveal some arbitrariness in theory. As  a consequence, in the theory the assumptions adopted do not seem to be completely justified  from a \emph{rational} point of view
 (see \cite{Acta} in  the  classical case). 
Because of these reasons, Liu, M\"uller and Ruggeri (LMR) \cite{LMR} (see also \cite{RET}) explored the possibility of having a new theory that starts with a few natural assumptions and that uses only universal principles. This new tentative of Extended Thermodynamics, called sometimes Rational Extended Thermodynamics (RET) [12], was motivated by the kinetic theory at the mesoscopic level and is based on the following general assumptions: the field is formed by $14$ unknown 
and in correspondence, we have a system of $14$ balance laws formed by the conservation laws of particle numbers,  energy-momentum 
\begin{align}\label{first2}
\partial_\alpha V^\alpha =0, \qquad
\partial_\alpha T^{\alpha\beta}=0,
\end{align}
and 
\begin{equation}\label{AtripleNE}
\partial_\alpha A^{\alpha<\beta \gamma>} = I^{<\beta \gamma>}.
\end{equation}
The  $A^{\alpha<\beta \gamma>}$ is the deviatoric part concerning the last two indices of a triple tensor that is symmetric with respect to all indices  $A^{\alpha \beta \gamma}$. The Greek index runs from $0$ to $3$, $\partial_\alpha =  \partial/\partial x^\alpha$, and we adopt the summation convection, i.e., we take summation over repeated indexes from $0$ to $3$.
As it is well known, we can decompose $V^\alpha$ and $T^{\alpha \beta}$ using physical variables:
\begin{align}
& V^\alpha =  n  m U^\alpha, ~\text{is the particle number four-vector},  \label{Va} \\
& T^{\alpha \beta} =  t^{<\alpha \beta>_3}  + (p+ \pi ) h^{\alpha \beta} +
\frac{2}{c^2} \, U^{(\alpha } q^{\beta)} + \frac{e}{c^2} \, U^{\alpha } U^{\beta}  \,\, \text{is the energy-momentum tensor},\label{Tab}
\end{align}
where $\rho = n m$, $n$ is the  particle number, $m$  is the mass in the rest frame, $U^\alpha$ is the four-velocity vector, $g^{\alpha\beta}$ is the metric tensor with signature $(+,-,-,-)$, 
$\, h^{\alpha \beta} = \ua \ub/c^2 - g^{\alpha\beta} $  is the projector tensor, $p$ the equilibrium pressure, $\pi$ the dynamical pressure, $c$ the light speed  and 
\begin{align}
& e= T^{\alpha\beta} U_\alpha U_\beta = \rho(c^2+\varepsilon) \quad \text{is energy}, \quad  (\varepsilon \,\  \text{is internal energy density}), \label{ee}\\
&     t^{<\alpha \beta>_3} = T^{\mu\nu} \left(h^\alpha_\mu h^\beta_\nu - \frac{1}{3}h^{\alpha\beta}h_{\mu\nu}\right) ~\text{ is the deviatoric shear viscous stress tensor},  \nonumber\\
&    q^\alpha= -h^\alpha_\mu U_\nu T^{\mu \nu} 
\,\, \text{is the heat flux four-vector},  \nonumber
\end{align}
with the  constraints:
\begin{equation} \label{C-1}
\ua U_\alpha= c^2, \quad h^{\alpha\beta} U_\beta =0, \quad t^{<\alpha \beta>_3}  U_\beta = 0, \quad g_{\alpha\beta} t^{<\alpha \beta>_3} =0, \quad q^\alpha U_\alpha = 0.  
\end{equation}
The $14$ unknown of the system \eqref{first2}, \eqref{AtripleNE} are ${\bf u} \equiv (V^\alpha, T^{\alpha \beta})$, or equivalently  the independent variables ${\bf u} \equiv (\rho, e, U^\alpha, q^\alpha, t^{<\alpha \beta>_3}, \pi)$ taking into account the constraints \eqref{C-1}. The RET aims to close the system by giving the expressions of the triple tensor $A^{\alpha<\beta \gamma>}$ and of the production tensor $I^{<\beta \gamma>}$ in terms of the field ${\bf u}$ using universal principles as entropy and relativity principles and convexity of entropy.

The kinetic theory motivates the theory, for which the triple tensor is a moment of the distribution function. The same assumptions were given successively by Pennisi and Ruggeri \cite{PR} and by Arima, Carrisi, Pennisi, and Ruggeri  \cite{ACPR}  for polyatomic gases where the triple tensor is a more complex moment of an appropriate new extended distribution function. Both theories converge formally in the parabolic limit to the Eckart equations using the so-called Maxwellian iteration (see \cite{LMR,CPR,ACPR}).

The Maxwellian iteration was introduced first by Ikenberry and Truesdell \cite{IT}, and it is substantially composed of (i) an identification of the relaxation times and (ii)  a formal power expansion of the solution in terms of the relaxation times: a sort of Chapman-Enskog procedure at the macroscopic level.
In general, the first iterates 
are obtained from the right-hand sides of balance laws by putting the ``zeroth" iterates ---
equilibrium values ---  into the left-hand sides. The second iterates
are obtained from the right-hand sides by putting the first iterates into the left-hand sides, and 
so on.   Substantially, we imagine the non-equilibrium variables as power expansion with respect to a relaxation time $\tau$ and the equation \eqref{AtripleNE} like
\begin{equation*}
\partial_\alpha A^{\alpha<\beta \gamma>} = \frac{1}{\tau}I^{<\beta \gamma>}.
\end{equation*}
Then  at zeroth  level $I_{(0)}^{<\beta \gamma>} =0$, while
\begin{equation}\label{Atriple0}
\partial_\alpha A_E^{\alpha<\beta \gamma>} = I_{(1)}^{<\beta \gamma>}.
\end{equation}
This means that the Eckart equations    can be obtained as the first iterate parabolic limit of the RET   substituting in \eqref{AtripleNE} the equilibrium part of the triple tensor   
and taking into account the equilibrium expression of the conservation laws \eqref{first2}: 
\begin{equation}\label{equil}
\partial_\alpha V^\alpha = 0, \qquad \partial_\alpha T_E^{\alpha \beta}=0.
\end{equation}
This technique was successful both in classical and relativistic frameworks to obtain the parabolic counterpart of RET theories (see for more details \cite{RET,Newbook}).

All previous theories of RET were motivated by the Kinetic Theory (KT) at the mesoscopic scale and are in perfect agreement with the closure of moments of a distribution function. The limit of the theory, as the KT,  is that the range of validity requires the gas to be rarefied. 

Also, for possible application to nuclear physics, the question is if it is possible to construct a pure macroscopic model of RET  for dense gas when the kinetic theory is unavailable.

This paper tries to give the first answer to this problem. In particular, we want to study under which conditions a general system of $14$ balance laws \eqref{first2}, \eqref{AtripleNE} give as the first iterate in the Maxwellian iteration the  Eckart equations without requiring the structure of moments of a distribution function for the tensors in the balance equation system. 

As it is well known, any physical hyperbolic theory when "relaxation times" are negligible needs to obtain the corresponding parabolic model as in the classical framework. Therefore this is a mandatory test for constructing a model for a generic kind of gas of hyperbolic type with a solid physical background.

We want to prove that the requirement that at first iterate equation \eqref{AtripleNE} converges to the Eckart equations is not automatically verified and implies a constraint between the two equilibrium functions present in the triple tensor evaluated in equilibrium. Moreover, the production tensor
$ I^{<\beta \gamma>}$ is completely determined from this request. The symmetric form induced by the entropy principle using the symmetrization procedure by Ruggeri and Strumia \cite{RS} permits us to prove that a general RET theory satisfies the compatibility condition and, therefore,  a general divergence causal theory always converges in the parabolic limit to the Eckart equations.

\section{Eckart  equations} 
A pioneer of Relativistic Thermodynamics is Carl Eckart \cite{Echart}  who, as early as 1940, established the
Thermodynamical of Irreversible Processes, a theory now known by the acronym TIP. Eckart's theory is
an important step away from equilibria towards non-equilibrium processes. 
It provides a counterpart
of the Navier–Stokes equations for the viscous stress tensor and a generalization of Fourier's law of heat
conduction. The latter permits a heat flux to be generated by an acceleration or a temperature gradient
to be equilibrated by a gravitational field. However, Eckart's theories have one drawback: they lead to
parabolic equations and thus predict infinite pulse speeds in contrast
with relativity.

In this section,  as it is preliminary for our results, we review the Eckart procedure. The field equations in TIP are the conservation of particle-particle flux and the conservation of energy-momentum \eqref{first2}.
We split the equations in the above system into their temporal and spatial components,
\begin{align} 
& \partial_{\alpha} (\rho U^{\alpha})= 0 \, , \label{mass} \\
& U_{\beta}\partial_{\alpha} T^{\alpha\beta}= 0  \label{energy}\, , \\
& h_{\beta\gamma}\partial_{\alpha} T^{\alpha\beta}= 0  \label{momentum}.
\end{align}
The  \eqref{mass} becomes the conservation of particles number:
\begin{equation}\label{massa}
\dot{\rho} + \rho \, \partial_\alpha U^\alpha = 0,
\end{equation}
where for any tensorial  function $f(x^\beta) $, we introduce the material derivative with respect to the proper time as:
\begin{equation}\label{Dot}
\dot{f} =  U^\alpha \partial_\alpha f.
\end{equation}
Taking into account that  from \eqref{Tab}, \eqref{ee} and \eqref{C-1} we have:
\begin{equation}\label{pass1}
T^{\alpha\beta}U_\beta = e  U^\alpha + q^\alpha = \rho \, (c^2 + \varepsilon) U^\alpha + q^\alpha,
\end{equation}
then \eqref{energy} becomes:
\begin{equation}\label{pass2}
0 = U_{\beta}\partial_{\alpha} T^{\alpha\beta}= \partial_{\alpha} \left(T^{\alpha\beta}U_\beta \right) -T^{\alpha\beta} \partial_\alpha U_\beta.
\end{equation}
Taking into account  \eqref{pass1}, \eqref{mass} and \eqref{Dot},  the \eqref{pass2} become the evolution equation for the internal energy:
\begin{equation} \label{energia}
\rho \dot{\varepsilon} + \partial_{\alpha} q^\alpha - \frac{1}{c^2} q^\alpha\dot{U}_\alpha -P^{\alpha \beta} \partial_\alpha U_\beta = 0,
\end{equation}
with
\begin{equation*}
P^{\alpha \beta} = t^{<\alpha \beta>_3} -(p +\pi) g^{\alpha\beta},
\end{equation*}
that have the meaning of the total relativistic stress tensor.

Taking into account that $\varepsilon$ is a function   of $\rho, T$, the equation of energy \eqref{energia} can be written as an equation for the temperature (taking into account \eqref{massa}):
\begin{equation} \label{enuno}
\rho c_V\dot{T} + \partial_{\alpha} q^\alpha - \frac{1}{c^2} q^\alpha\dot{U}_\alpha -P^{\alpha \beta} \partial_\alpha U_\beta -\rho^2 \varepsilon_\rho \partial_\alpha U^\alpha = 0.
\end{equation}
In relativity, the Gibbs equation is the same as the classical one:
\begin{equation}\label{Gibbs}
T d S = d\varepsilon -\frac{p}{\rho^2} d\rho \quad \longleftrightarrow \quad  T d S = \frac{de}{\rho} -\frac{e+p}{\rho^2} d\rho,
\end{equation}
where $S$ is the equilibrium entropy.
Taking into account that from the integrability of the Gibbs equation \eqref{Gibbs}$_2$, we have
\begin{equation*}
e_\rho = \frac{e+p-T p_T}{\rho},
\end{equation*}
the \eqref{enuno}  can  be rewritten as
\begin{align}\label{Temperatura2}
e_T \dot{T} + \partial_{\alpha} q^\alpha - \frac{1}{c^2} q^\alpha\dot{U}_\alpha -t^{{<\alpha \beta>}_3} \partial_\alpha U_\beta +T p_T \, \partial_{\alpha} \ua =0,
\end{align}
where we put 
\begin{equation*}
c_V = \varepsilon_T=\left(\frac{\partial \varepsilon}{\partial T}\right)_\rho, \qquad \varepsilon_\rho = \left(\frac{\partial \varepsilon}{\partial \rho}\right)_T, \quad e_T = \left(\frac{\partial e}{\partial T}\right)_\rho, \qquad e_\rho = \left(\frac{\partial e}{\partial \rho}\right)_T.
\end{equation*}

The equation \eqref{momentum} represents the momentum equation that can be put in a form similar to the previous ones, but it is not needed at the moment.

From \eqref{Gibbs} we have:
\begin{equation*}
T \dot{S} = \dot{\varepsilon} -\frac{p}{\rho^2} \dot{\rho} \, .
\end{equation*}
Therefore taking into account \eqref{massa} and \eqref{energia} we have
\begin{align*} 
\rho \dot{S}+ \partial_\alpha\left(\frac{q^\alpha}{T} \right)= -\frac{q^\alpha}{T^2}\left(\partial_\alpha T - \frac{T}{c^2}\dot{U}_\alpha\right)+\frac{1}{T}\left(t^{<\alpha \beta>_3}\partial_\alpha U_\beta -\pi \partial_\alpha U^\alpha\right),
\end{align*}
that is equivalent to the entropy law
\begin{equation*}
\partial_\alpha h^\alpha = \sigma,
\end{equation*}
with 
\begin{align*}
h^\alpha = \rho S U^\alpha + \frac{q^\alpha}{T}, \qquad \sigma= -\frac{q^\alpha}{T^2}\left(\partial_\alpha T - \frac{T}{c^2}\dot{U}_\alpha\right)+\frac{1}{T}\left(t^{<\alpha \beta>_3}\partial_\alpha U_\beta -\pi \partial_\alpha U^\alpha\right),
\end{align*}
As   $\sigma$ is the entropy production and must be positive, in TIP, it is assumed that the proportionality between fluxes and forces is. However, we need to take into account the constraints \eqref{C-1}, and therefore Eckart \cite{Echart} obtained with this procedure  the relativistic counterpart of Navier-Stokes-Fourier equations:
\begin{align}\label{8b}
\begin{split}
& \pi= -\nu \, \partial_{\alpha} U^{\alpha}\, , \\
& q_{\beta}=-\chi \, h ^\alpha_\beta\left(\partial_\alpha T - \frac{T}{c^2}\dot{U}_\alpha\right) \, ,\\
&	t_{<\beta \delta>_3} =2 \mu  \, h^\alpha_\beta\, h^\mu_\delta\partial_{<\alpha} U_{\mu>_3} \, ,
\end{split}
\end{align}
where $\nu$, $\chi$, and $\mu$ are the bulk viscosity, the heat conductivity, and the shear viscosity, respectively, that must be non-negative. The second term in \eqref{8b}$_2$ is the famous acceleration term that influences the heat flux in relativity.

\section{First order approximation in the Maxwellian Iteration}
Now we consider a hyperbolic system in which instead of \eqref{8b}, we have the balance equations \eqref{AtripleNE}, and we consider the first iteration  \eqref{Atriple0} omitting the index ${(1)}$ for simplicity.  Therefore we ask when  
\begin{equation}\label{Atriple}
\partial_\alpha A_E^{\alpha<\beta \gamma>} = I^{<\beta \gamma>},
\end{equation}
gives the Eckart equations \eqref{8b} in the first Maxwellian iteration.

As the left side of \eqref{Atriple} contains only equilibrium variables,  we need to represent the most general triple tensor in equilibrium. We consider a general triple tensor symmetric with respect to all indices also if this is not important as we consider only the deviatoric part, but this can be useful to compare later with the results of RET. 
The most general triple tensor with equilibrium variables can be written in the form:
\begin{equation}\label{A1triple}
A_E^{\alpha\beta\gamma} =  \bar{a} \, \ua \ub \ug +\bar{b} \, (\hab \ug + \hag \ub+ \hbg \ua),
\end{equation}
where $\bar{a}, \bar{b}$ functions of $(\rho, T)$.
While the right  side of \eqref{Atriple} contains only non-equilibrium variables vanishing in equilibrium, then
\begin{equation*}
	I^{\beta\gamma} =  a_1 \, \left(\ub \qg + \ug \qb\right) + a_2 \, \tbg + a_3 \, \pi \ub \ug + a_4 \,\pi\gab,
\end{equation*}
with $a_1, a_2, a_3, a_4$ functions of $(\rho,T)$.

We recall that,  
for any symmetric tensor $M^{\alpha \beta}$, we can define its traceless part  $M^{< \alpha \beta >}$ and its 3-dimensional traceless part  $M^{< \alpha \beta >_3}$ that is the traceless part of its projection in the 3-dimensional space orthogonal to $U^\alpha$	as follows
\begin{align}
		&  M^{< \alpha \beta >} = \left( g_\mu^{\alpha} \,  g_\nu^{ \beta} - \, \frac{1}{4} \, g^{\alpha \beta} g_{\mu \nu} \right) \, M^{\mu \nu} = M^{\alpha \beta}  - \, \frac{1}{4} \,  g_{\mu \nu}  \, M^{\mu \nu} g^{\alpha \beta} \, , \label{devia1}\\
		& M^{< \alpha \beta >_3} = \left( h_\mu^{\alpha} \,  h_\nu^{ \beta} - \, \frac{1}{3} \, h^{\alpha \beta} h_{\mu \nu} \right) \, M^{\mu \nu} \nonumber \, ,
\end{align}
which are different except in the case in which 
$M^{\mu \nu} U_\mu=0$, and $M^{\mu \nu} g_{\mu \nu}=0 $. 
Taking into account the deviatoric definition \eqref{devia1} we have:
\begin{align}
&	A^{\alpha<\beta\gamma>} =  a \, \ua \left(\hbg +\frac{3}{c^2}  \ub \ug  \right)+ b \, \left( \hag \ub+ \hab \ug \right), \label{aaa}\\
& 	I^{<\beta\gamma>} =  a_1 \,  \left(\ub \qg + \ug \qb\right) + a_2 \, \tbg + a_3 \, \pi \left(\ub \ug -\frac{c^2}{4}\gbg\right), \label{iii}
\end{align}
where 
\begin{equation}\label{abab}
	a = \frac{1}{4}\left(\bar{b}+\bar{a} c^2\right), \qquad b = \bar{b}.
\end{equation}
Of course, now the tensor $A^{\alpha<\beta\gamma>}$ is no  more symmetric in all indices.
The coefficients $a,b,a_1,a_2,a_3$ depending only on $(\rho, T)$ can be determined such that \eqref{Atriple} are identical to the Eckart equations \eqref{8b} in the first approximation of Maxwellian iteration.
We   prove the following 
\begin{theorem}\label{teorema1}
Necessary and sufficient condition such that   the divergence form \eqref{Atriple} in the approximation of first order Maxwellian iteration converges to the Eckart equations \eqref{8b} is that the coefficients $a$, $b$  satisfy the relation
\begin{align}\label{ab0}
	a= \frac{1}{4} \left\{-b+ \left(e+p -T p_T\right)\frac{b_\rho}{p_\rho}+ T \, b_T
	\right\}.
\end{align}
	Moreover there exist the following relations between the coefficients $a_1, a_2, a_3$ and the phenomenological coefficients $\nu, \mu, \chi$:
	\begin{align}\label{bellino}
	\begin{split}
			&	a_1= \frac{1}{p_\rho\, \chi}\left(b_\rho \, p_T -b_T \, p_\rho \right), \\
		& 	a_2 = -\frac{b}{\mu}, \\
	&	a_3 =-\frac{4}{c^2 \nu}\left[a+\frac{2}{3}b-a_\rho \, \rho - \frac{a_T}{e_T}T p_T\right].
	\end{split}
	\end{align}
	\end{theorem}

\textbf{Proof}:	Proceeding in the same way as in the  LMR paper \cite{LMR} or in CPR \cite{CPR}, we  obtains the Eckart equations \eqref{8b} from \eqref{Atriple} trough the following independents projections:
\begin{align}\label{AtripleE}
\begin{split}
	&\Ub \Ug	\partial_\alpha A_E^{\alpha<\beta \gamma>} = I^{<\beta \gamma>} \Ub \Ug, \\
&\Hbd \Ug	\partial_\alpha A_E^{\alpha<\beta \gamma>} = I^{<\beta \gamma>} \Hbd \Ug, \\
&\left(\Hbm \Hgn -\frac{1}{3} \Hmn \Hbg\right)	\partial_\alpha A_E^{\alpha<\beta \gamma>} = I^{<\beta \gamma>} \left(\Hbm \Hgn -\frac{1}{3} \Hmn \Hbg\right) ,
\end{split} 
\end{align}
The first  of \eqref{AtripleE} can be rewritten:
\begin{equation*}
\partial_\alpha\left(\Ub \Ug A_E^{\alpha<\beta \gamma>}\right) -A_E^{\alpha<\beta \gamma>} \partial_\alpha\left(\Ub \Ug\right)= I^{<\beta \gamma>} \Ub \Ug
\end{equation*}
that implies
\begin{equation}\label{primaeq}
	\dot{a}+\left(a +\frac{2}{3}b\right) \partial_\alpha \ua = \frac{1}{4}c^2 a_3\,  \pi.
\end{equation}
Now $\dot{a} = a_\rho \, \dot{\rho} + a_T \, \dot{T}$ and recalling that we consider the first iterate, we need to use the 
equilibrium values of the conservation laws \eqref{equil} and in particular for $\dot{\rho}$ and $\dot{T}$. For this aim, we can eliminate these quantities using 
 the number of particle equation \eqref{massa} and  the equation for the temperature \eqref{Temperatura2} in equilibrium 
\begin{equation}\label{tempEqu}
		e_T \dot{T}  +T p_T\, \partial_{\alpha} \ua =0.
\end{equation}
Then the  \eqref{primaeq}  is equivalent to the Eckart  Eq. \eqref{8b}$_1$ provided \eqref{bellino}$_3$ is satisfied.

The second   of \eqref{AtripleE} can be rewritten:
\begin{equation*}
	\Hbd \left\{\partial_\alpha  \left(\Ug	 A_E^{\alpha<\beta \gamma>}\right) -  A_E^{\alpha<\beta \gamma>}\partial_\alpha  \Ug\right\} = I^{<\beta \gamma>} \Hbd \Ug
\end{equation*}
that is equivalent to
\begin{equation}\label{seconda}
\Hbd \left\{-c^2 \partial^\beta b +(4a +b) \dot{U}^\beta \right\} = -c^2 a_1 \, q_\delta.
\end{equation}
The momentum equation \eqref{momentum} evaluated in equilibrium gives
\begin{equation}\label{momeq}
	\frac{e+p}{c^2 }\dot{U}^\beta +h^{\mu\beta} \, \partial_\mu p =0.
\end{equation}

Substituting \eqref{momeq} in \eqref{seconda} and in \eqref{8b}$_2$ and requiring that \eqref{seconda} is the same as Eckart \eqref{8b}$_2$ we have two conditions as $\partial^\beta \rho$ and $\partial^\beta T$ are independent:
\begin{align*}
&(e+p)\, 	b_\rho -(4a + b) \, p_\rho  = \chi a_1 T p_\rho,\\
& (e+p)\,	b_T -(4a + b)  \, p_T = - \chi a_1 (e+p - T \, p_T).\\
\end{align*}
From which we have the constrain  between $a$ and $b$ given in \eqref{ab0} 
and the relation between $a_1$ and $\chi$ given in \eqref{bellino}$_1$.
Finally, it is simple to verify that \eqref{AtripleE} coincides with the last equation of Eckart \eqref{8b}$_3$ if \eqref{bellino}$_2$ is true.

\begin{remark}
It is important to observe that the Eckart system formed by the equations \eqref{first2} and \eqref{8b}, in general do not coincide with the system \eqref{first2} and \eqref{Atriple} (with coefficients given in the theorem 1). This is because we proved the identity \eqref{8b} with \eqref{Atriple} using the equilibrium conservation laws \eqref{equil} according to the Maxwellian iteration.  
We ask now when the two systems are perfectly equal. 
In the proof of theorem 1, we have used first \eqref{tempEqu} instead of \eqref{Temperatura2}. Then to have perfect coincidence between \eqref{primaeq} and \eqref{8b}$_1$ we need $a_T =0$. Moreover to be identically \eqref{seconda} with \eqref{8b}$_2$ we need
\begin{equation*}
\frac{1}{a_1}\partial_\beta b = -\chi \partial_\beta T, \quad  \frac{1}{a_1}(4a +  b) =-\chi T,
\end{equation*}
i.e.
\begin{equation*}
a= c_1, \quad b= c_2 T - 4 c_1, \quad a_1= -\frac{c_2}{\chi  }
\end{equation*}
that except  for inessential constants $c_1, c_2$ coincide with the form written first by Geroch and Lindblom \cite{GL} in which $a=0, b=T$. In this case \eqref{ab0} it is trivial satisfied while \eqref{bellino} becomes
\begin{equation*}
a_1 = -\frac{1}{\chi}, \quad a_2 = -\frac{T}{\mu}, \quad a_3 = - \frac{8 T}{3 c^2 \nu},\end{equation*}
and therefore
\begin{align*}
&	A_E^{\alpha<\beta\gamma>} = T \, \left( \hag \ub+ \hab \ug \right), \\
& 	I^{<\beta\gamma>} =  -\frac{1}{\chi} \,  \left(\ub \qg + \ug \qb\right)  -\frac{T}{\mu} \, \tbg  -\frac{8 T}{3 c^2 \nu}\, \pi \left(\ub \ug -\frac{c^2}{4}\gbg\right). 
\end{align*}
\end{remark}

\section{Symmetric Hyperbolic Systems of Divergence type and Eckart limit}
Now we want to prove that any causal  theory of divergence type compatible with an entropy principle as the one of RET automatically satisfies the compatibility condition \eqref{ab0}, and therefore, it converges in the first Maxwellian iteration to the Eckart system.
\begin{theorem}\label{teorema2}
    If the system of balance equations
\begin{equation}\label{1}
\partial_\alpha V^\alpha =  0   \quad , \quad
\partial_\alpha T^{\alpha \beta} =  0 \quad , \quad \partial_\alpha A^{\alpha < \beta \gamma >} =  I^{ < \beta \gamma >} \, ,
\end{equation}
is compatible with  the entropy principle, i.e., there exists an entropy 4-vector $h^\alpha$ such that   for any thermodynamical processes,
\begin{equation*}
\partial_\alpha h^\alpha = \sigma \geq 0,
\end{equation*}
then the system assumes a symmetric form using the main field and automatically satisfies the compatibility condition \eqref{ab0}.
\end{theorem}
\textbf{Proof}:	
Using the symmetrization technique given by Ruggeri and Strumia \cite{RS},   there exists a 4-vector potential 
$h'^\alpha$ and a \emph{main field} $\mathbf{u}^{\prime}\equiv(\lambda, \lambda_\beta, \Sigma_{\beta\gamma})\, \,$
($\Sigma_{\beta\gamma} $ is a symmetric and deviatoric tensor)  such that :
\begin{align}
& V^\alpha =\frac{\partial h'^\alpha}{\partial \lambda}, \quad T^{\alpha\beta} =\frac{\partial h'^\alpha}{\partial \lambda_\beta},  \quad A^{\alpha <\beta\gamma>} =\frac{\partial h'^\alpha}{\partial \Sigma_{\beta\gamma}} -\frac{1}{4} g^{\beta\gamma} g_{\mu\nu}\frac{\partial h'^\alpha}{\partial \Sigma_{\mu\nu}}, \label{mainf}\\
& h'^\alpha= \lambda V^\alpha + \lambda_\beta T^{\alpha \beta} +\Sigma_{\beta\gamma} A^{\alpha\beta \gamma} - h^\alpha \nonumber \\
&\sigma = \Sigma_{\beta \gamma} I^{\beta\gamma} \geq 0, \nonumber
\end{align}
and the system \eqref{first2} and  \eqref{AtripleNE} assumes the symmetric Godunov form:
\begin{equation}\label{sform} 
\partial_{\alpha} \left(\frac{\partial h^{\prime\alpha}}{\partial {\bf u}^\prime}\right) = {\bf f }, \qquad  \leftrightarrow \qquad 
 \frac{\partial^2 h^{\prime\alpha}}{\partial \mathbf{u}^\prime   \partial  \mathbf{u}^\prime  } \partial_\alpha {\bf u}^\prime = {\bf f }.
\end{equation}

By the representation theorem, we have the expression (A.2) of \cite{LMR}:
\begin{equation}\label{z1}
h'^\alpha = \sum_{A= 0}^{3} \gamma_A \lambda^{(A) \alpha} \, ,
\end{equation}
with  $\lambda^{(A) \alpha}= (\Sigma^A)^{\alpha \beta}  \lambda_\beta,$ where  the power $A=0,1,2,3$ for the  Hamilton-Cayley theorem (($\Sigma^0)^{\alpha \beta}=  g^{\alpha\beta}$) and the coefficients $\gamma_A$ may be functions of all scalars $\lambda, \, G_A = \lambda^{(A) \beta} \lambda_\beta$ and $Q_i = (\Sigma^{i+1})^{\beta}_{\,\, \beta}$, $\, (i=1,2)$.

Imposing the symmetry of  $T^{\alpha \beta}$ and $A^{\alpha  \beta \gamma }$ we still obtain  eq. (A.3)  of \cite{LMR}. As $\Sigma^{\alpha \beta}=0$ in equilibrium we have up to the second order \cite{PR}
\begin{align}\label{z2}
\begin{split}
&    \gamma_2 = \Gamma_2 \, , \\
&    \gamma_1 = \Gamma_1 + \frac{\partial  \Gamma_2}{\partial G_0} G_1 \, ,   \\
&    \gamma_0 = \Gamma_0 + \frac{\partial  \Gamma_1}{\partial G_0} G_1 +  \frac{1}{2} \frac{\partial^2  \Gamma_2}{\partial G_0^2} G_1^2 + \frac{\partial  \Gamma_2}{\partial G_0} G_2 + \frac{1}{4} \Gamma_2 \Sigma^{\mu \nu} \Sigma_{\mu \nu} \, ,    
\end{split}
\end{align}
where $\Gamma_0$, $\Gamma_1$, $\Gamma_2$ are arbitrary functions of $\lambda$, $G_0= \lambda^\alpha \lambda_\alpha$.  
Consequently, at every order concerning equilibrium, a new arbitrary function appears, that is, $\Gamma_0$ at the order zero, $\Gamma_1$ at the order one and $\Gamma_2$ at the order two. \\

Independently of the order of the truncation as a consequence of \eqref{z1} and  \eqref{z2}, taking into account that in equilibrium $\Sigma^{\beta \gamma}=0$,  we have the following expression for the triple tensor in equilibrium (A.6) of \cite{LMR}:
\begin{equation}\label{A1tripleLMR}
A_E^{\alpha\beta\gamma} =  \frac{\partial \Gamma_1}{\partial G_0} \, \la \lb \llg +\frac{1}{2} \Gamma_1\, (\gab \llg + \gag \lb+ \gbg \la),
\end{equation}
where the quantities to the right side are evaluated at equilibrium.
For the   equilibrium system \eqref{equil}  the \emph{main field} was calculated by Ruggeri and Strumia   \cite{RS} as
\begin{equation}\label{MFE}
\lambda = -\frac{g_r}{T}, \quad   \lambda^\beta = \frac{\ub}{T},  \qquad \left(G_0= \frac{c^2}{T^2}, \quad g_r= \frac{e+p}{\rho}- T S\right).
\end{equation}
Inserting \eqref{MFE} into \eqref{A1tripleLMR} and comparing with \eqref{A1triple} we obtain
\begin{equation*}
\bar{a} =\frac{3  \Gamma_1}{2 c^2
	T}+ \frac{\partial \Gamma_1}{\partial G_0} \frac{1}{T^3}, \qquad \bar{b} = -\frac{\Gamma_1}{2  T},
\end{equation*}
then from \eqref{abab}
\begin{equation}\label{554}
a = \frac{1}{4}\left(\frac{  \Gamma_1}{ 
	T}+ \frac{\partial \Gamma_1}{\partial G_0} \frac{c^2}{T^3}\right), \quad b = -\frac{\Gamma_1}{2  T}.
\end{equation}
Now we  pass from the variables $(\lambda,G_0)$ to $(\rho,T)$. From \eqref{MFE} and taking into account the Gibbs equation \eqref{Gibbs}$_2$, we have
\begin{equation*}
d \lambda = \frac{dT (e+p-p_T
	T)-p_\rho T d\rho }{\rho 
	T^2}, \qquad dG_0= -\frac{2 c^2 dT}{T^3},
\end{equation*}
then
\begin{equation}\label{cinque}
\frac{\partial \Gamma_1}{\partial \lambda}  = -\frac{\partial \Gamma_1}{\partial \rho} \frac{T \rho}{p_\rho}, \quad
 \frac{\partial \Gamma_1}{\partial G_0}  = -\frac{T^2}{2 c^2 p_\rho}\left\{
 \frac{\partial \Gamma_1}{\partial \rho} \left(e + p - T p_T \right) +\frac{\partial \Gamma_1}{\partial T} p_\rho T \right\}.
\end{equation}
Inserting the second of \eqref{cinque} into \eqref{554}
\begin{equation}\label{last}
a = \frac{1}{4}\left\{\frac{  \Gamma_1}{ 
	T}  -\frac{1}{2 T p_\rho}\left(
\frac{\partial \Gamma_1}{\partial \rho} \left(e + p - T p_T \right) +\frac{\partial \Gamma_1}{\partial T} p_\rho T \right)
\right\}, \quad b = -\frac{\Gamma_1}{2  T}.
\end{equation}
Then, if we insert in the first \eqref{last}  $\Gamma_1$ in term of  $b$ from the second relation of \eqref{last}, we obtain exactly the expression for $a$ given by \eqref{ab0}. 

\emph{Thus  it is proved that any causal  system  \eqref{1} that is  symmetric in the form \eqref{mainf} converges in the first  Maxwellian Iteration to the Eckart system!}

As examples that confirm the previous result, we present the particular case of RET of monatomic and polyatomic gases.
\subsection{RET of Monatomic Gases}
The Liu M\"uller Ruggeri theory \cite{LMR} is devoted to monatomic gas and is perfectly compatible with the relativistic theory in which $V^\alpha, T^{\alpha \beta}, A^{\alpha\beta\gamma}$ are the following moments of the distribution function $f(x^\alpha,  p^\beta)$:
\begin{align*} 
\begin{split}
& V^\alpha  = m c \int_{\R^{3}} f p^\alpha   \, d \boldsymbol{P}  \, , \quad
  T^{\alpha \beta}  = c \int_{\R^{3}}  f p^\alpha p^\beta \, d \boldsymbol{P}  \, , \quad
 A^{\alpha \beta \gamma  } = \frac{c}{m} \int_{\R^{3}}
 f  \,   p^{\alpha} p^\beta p^{\gamma}    \, d \boldsymbol{P}  \, , 
\end{split}
\end{align*}
where $p^\alpha$ is the four-momentum and 
 \begin{equation*}
d \boldsymbol{P} =  \frac{dp^1 \, dp^2 \,
	dp^3}{p^0} .
\end{equation*}
In the  paper \cite{LMR}  (see also \cite{RET,Newbook}) the Eq. (7.12) coincides with \eqref{A1triple} with coefficients  (in the simple case of non-degenerate gases) given by the  Eq (7.13)$_4$ of LMR:
\begin{equation*}
	\bar{a}= \rho \left(1 + \frac{3 G}{\gamma}\right), \qquad \bar{b} = c^2 \frac{G\, \rho}{\gamma},
\end{equation*}
then from \eqref{abab}
\begin{equation*}
	a =\rho c^2 \left(\frac{1}{4} + \frac{ G}{\gamma}\right), \qquad {b} = c^2 \frac{G\, \rho}{\gamma},
\end{equation*}
where
\begin{equation*}
	G= \frac{K_3(\gamma)}{K_2(\gamma)}, \qquad \gamma = \frac{m c^2}{k_B T}.
\end{equation*}
 The $K_n(x)$ denotes the second-order Bessel functions and $k_B$ is the Boltzmann constant.
 Taking into account that in monatomic non-degenerate gas, we have (LMR (7.13)):
\begin{equation*}
	e = \rho c^2\left( G -\frac{1}{\gamma}\right), \quad p=\frac{\rho c^2}{\gamma}
\end{equation*}
and the well-known properties of the Bessel function for which
\begin{equation*}
	\frac{d G}{d \gamma } =-1- 5 \frac{ G}{\gamma} + G^2,
\end{equation*}
it easy to verify that the relation \eqref{ab0} is automatically satisfied. Moreover, \eqref{bellino} becomes:
\begin{align*}
	& a_1 = -\frac{p}{\chi T}(1+ 5 G - \gamma \, G^2), \\
	&a_2 = - \frac{
	p}{\mu}G, \\
&a_3 =-\frac{4p}{3c^2 \nu} 
	  \left(2
	G-\frac{3 (\gamma +G
		(6-\gamma  G))}{\gamma 
		(\gamma +G (5-\gamma 
		G))-1}\right) 
\end{align*}
which is equivalent to Eq. (7.22) of Liu M\"uller Ruggeri  paper \cite{LMR}   with the different symbols 	$B_1^\pi = - a_3 c^2/4,$
$B_4 = a_1, $ $ B_3 = a_2 $.

\subsection{RET of Polyatomic Gases}
Pennisi and Ruggeri (PR) \cite{PR} were the first to construct a model of polyatomic gas whose moments are given by
\begin{align} 
\begin{split}
& A^\alpha  = m c \inta f p^\alpha \phi(\mathcal{I}) \, d \mathcal{I} \, d \boldsymbol{P}  \, , \\
& A^{\alpha \beta}  = \frac{1}{mc} \inta f p^\alpha p^\beta (mc^2 + \II) \, \phi(\mathcal{I}) \, d \mathcal{I} \, d \boldsymbol{P}  \, , \\
& A^{\alpha \beta \gamma  } = \frac{1}{m^2 c} \int_{\R^{3}}
\int_0^{+\infty} f  \,   p^{\alpha} p^\beta p^{\gamma}  \, \Big( mc^2 + 2\II \Big) \, 
\phi(\mathcal{I}) \, d \mathcal{I} \, d \boldsymbol{P}  \, , 
\end{split}
\label{PS3}
\end{align}
where the distribution function $f(x^\alpha, p^\beta,\mathcal{I})$ depends on the extra variable $\mathcal{I}$, similar to the classical one (see \cite{Newbook} and references therein) that has the physical meaning of the \emph{ molecular internal energy of internal modes} in order to take into account the exchange of energy due to the rotation and vibration of a molecule, and $\phi(\mathcal{I})$ is the state density of the internal mode.

 In this case, the deviatoric triple tensor in equilibrium is given by Eq. (48) of PR paper with coefficients $A_1^0$ and $A_{11}^0$  that are linear in  $\rho$ and complex function of the temperature (see Eqs (49)-(50) of PR paper). In this case
$\bar{a} =A_1^0$ and $\bar{b}=A_{11}^0 $ and therefore 
\begin{equation}\label{459}
a =\frac{1}{4}\left( c^2 A_1^0+A_{11}^0\right), \qquad {b} = A_{11}^0.
\end{equation}
Taking into account that in polyatomic gas, we have (PR (40)):
\begin{equation}\label{eep}
e = \rho c^2 \omega(\gamma), \quad p=\frac{\rho c^2}{\gamma},
\end{equation}
  the compatibility condition \eqref{ab0} become
\begin{equation}\label{aCPR}
a=\frac{1}{4}\left\{ b\left(\frac{e}{p}-1
\right)  + T \, b_T
\right\}
\end{equation}
that taking into account \eqref{459}, 
corresponds to the Eq. (A2) of the Carrisi, Pennisi, and Ruggeri  (CPR) paper \cite{CPR}, that considered the Maxwellian iteration of the Pennisi - Ruggeri model of polyatomic relativistic gas \cite{PR}.
 In the CPR paper, the identity (A2)  was a difficult proof due to the particular property of some integral of the distribution function. Instead this is a general property thanks to  Theorem \ref{teorema2}.

Taking into account the linearity  of $a, b, e$ and $p$ with respect to $\rho$, the equations  \eqref{bellino}$_1$ becomes
\[a_1 = \frac{1}{\chi T}(b - T \, b_T)\] 
and eliminating $b_T$ from \eqref{aCPR} we have
\begin{equation}\label{a1CPR}
a_1 = -\frac{1}{\chi T }\left(4 a - b \frac{e}{p}\right).
\end{equation}
while from the remaining equations of \eqref{bellino} we have
\begin{equation}\label{a2a3CPR}
a_2= - \frac{b}{\mu}, \qquad a_3 = -\frac{4}{c^2 \nu}\left(\frac{2}{3}b - \frac{a_T}{e_T}p\right).
\end{equation}
The equations \eqref{a1CPR} and \eqref{a2a3CPR} taking into account \eqref{459} coincide with the equations (25) of the CPR paper \cite{CPR}.

Very recently, Arima, Carrisi, Pennisi, and Ruggeri   \cite{ACPR}, starting from an idea of Pennisi \cite{Pen}, proposed a more realistic model of polyatomic gas in which the Eq. \eqref{PS3} is substituted with the moment
\begin{equation*} A^{\alpha \beta \gamma  } = \frac{1}{m^3 c^2} \int_{\R^{3}}
\int_0^{+\infty} f  \,   p^{\alpha} p^\beta p^{\gamma}  \, \left( mc^2 + \II \right)^2 \, 
\phi(\mathcal{I}) \, d \mathcal{I} \, d \boldsymbol{P} .
\end{equation*}
In the present case, all equilibrium coefficients depend only on one scalar function of the temperature $\omega(\gamma)$, in particular
\begin{equation*}
a= \frac{1}{4} c^2 \rho 
\left(\frac{1}{\gamma
	^2}+\frac{\omega }{\gamma }+\omega
^2-\omega^\prime\right), \qquad b= \frac{c^2 \rho  (\gamma  \omega
	+1)}{\gamma ^2},
\end{equation*}  
with $e, p$ given in \eqref{eep}.
The model is also, in this case, a particular expression of the Theorem 1
and  again satisfies the compatibility condition \eqref{ab0} automatically.

\section{Macroscopic Relativistic Hyperbolic System}
Theorems \ref{teorema1} and \ref{teorema2} give light to the parabolic limit and permit to have  the expression of the production tensor. It remains open the triple tensor closure problem in non-equilibrium. 

In principle, in the main field variables, the problem is solved. In fact, for any choice of the functions, $\gamma_A$, the solution is given by \eqref{mainf} with \eqref{z1}. We need to add to the system \eqref{sform}  the requirement of convexity, i.e., that 
the matrix 
\begin{equation*}
\Ua \, \frac{\partial^2 h^{\prime\alpha}}{\partial \mathbf{u}^\prime   \partial \mathbf{u}^\prime  } \, , \qquad \text{must be positive negative},
\end{equation*}
 that it  is equivalent to the quadratic form:
\begin{align}\label{QQQ}
\begin{split}
Q = &\Ua\left\{\delta \lambda  \, \delta V^\alpha + \delta \lambda_\beta \, \delta T^{\alpha \beta} + \delta \Sigma_{\beta \gamma} \, \delta A^{\alpha<\beta \gamma>}\right\} = \\
&\Ua\left\{\delta \lambda  \, \delta \left(\frac{\partial h^{\prime\alpha}}{\partial \lambda}\right) + \delta \lambda_\beta \, \delta \left(\frac{\partial h^{\prime\alpha}}{\partial \lambda_\beta}\right) + \delta \Sigma_{\beta \gamma} \, \delta \left(\frac{\partial h^{\prime\alpha}}{\partial \Sigma_{\beta \gamma}}\right)\right\} < 0.
\end{split}
\end{align}
Unfortunately, the problem is the non-linear invertibility between the main field and  the physical variables.  For this reason, in the   RET theories, we usually  take   linear expressions for the non-equilibrium variables. In this approximation  we consider now  a macroscopic theory without any information at the mesoscopic scale, we want to prove that there are only two arbitrary functions of equilibrium variables $(\rho, T)$. In fact, from \eqref{z2} for what concerns the non-equilibrium triple tensor, we need only $\Gamma_0, \Gamma_1$ and $\Gamma_2$ that depend only on equilibrium variables. In these approximations, we have the expressions for $V^\alpha, T^{\alpha\beta} $ and $A^{\alpha\beta \mu}$ given in (A.14) and (A.16) of \cite{LMR}. In this case, the map between the physical variables and the main field components is linear, and the system can be closed regarding physical variables.

As was proved in \cite{LMR}  $\Gamma_0$  is obtained by the first two equations of \eqref{mainf} evaluated in equilibrium and we have:
\begin{equation*}
\Gamma_0 = -p,
\end{equation*}
while   we just proved in \eqref{554} that $\Gamma_1$ is related to $b$
\begin{equation*}
\Gamma_1 = - 2 T b,
\end{equation*}
that is in any way arbitrary together with $\Gamma_2$. Thanks to the Theorem \ref{teorema1}, the production tensor \eqref{iii} is entirely determined by the equations \eqref{bellino} provided that we know the thermal and the caloric equations of state and the phenomenological coefficients (heat conductivity, shear, and bulk viscosities).

Therefore any divergence theory \eqref{1} close to  equilibrium, compatible with an entropy principle, and having the Eckart system as limiting case when relaxation times are negligible,  have only two degrees of freedom, i.e., the equilibrium functions $b(\rho, T)$ and $\Gamma_2(\rho, T)$. The kinetic theory can give the precise value of these two functions both in monatomic  \cite{LMR} and a  polyatomic gas, \cite{PR,ACPR}. The only requirement that these two functions need is the requirement of convexity \eqref{QQQ} at least in the neighborhood of the equilibriums state.

An open problem is how we can choose  these two functions in a model where the kinetic theory is not applicable, like dense gases or solids.

\section{Classical limit}
We want to verify  that in the classical limit, Theorem 1 coincides with the similar result for a classical fluid given recently in \cite{AMR2023}.

We first recall that to obtain from relativistic balance laws the corresponding classical balance laws  one needs a linear combination as  reported in the book \cite{Newbook} on page 537:
The limit is obtained  when $c \rightarrow \infty$:
\begin{align} \label{Limiti}
\begin{split}
{\begin{array}{lll}
\partial_\alpha \, V^\alpha = 0 & \quad \rightarrow \qquad  \partial_t F + \partial_kF_k = 0,\\
& & \\
\partial_\alpha \, T^{\alpha \, i} = 0 & \quad \rightarrow \qquad  \partial_t F_i+ \partial_k F_{ki} =0,  \\
&& \\
2 \, \partial_\alpha \, \left(c \, T^{\alpha \, 0} \, - \, c^2 \, V^\alpha \right)= 0 & \quad \rightarrow \qquad \partial_t G_{ll} + \partial_k G_{llk} =0,  \\
& & \\
\partial_\alpha \, B^{\alpha \, \langle ij \rangle_3} =  I^{\langle ij \rangle_3}  & \quad \rightarrow \qquad   \partial_t H_{\langle ij \rangle} + \partial_k H_{k \langle ij \rangle} = P_{\langle ij \rangle},   \\
& & \\
\partial_\alpha \, \left(-4 \, B^{\alpha \, ij } \, g_{ij} -6 \, c \, T^{\alpha 0} + 3 c^2 V^\alpha \right) = - 4 I^{rs} g_{rs} & \quad \rightarrow \qquad   \partial_t H_{ll}   + \partial_k H_{kll}  = P_{ll},   \\
& & \\
2 \, \partial_\alpha \, \left( c \, B^{\alpha \, 0i} \, - \, c^2 \, T^{\alpha \, i} \right)=2 c  I^{0i} & \quad \rightarrow \qquad   \partial_t I_{lli} + \partial_k I_{llik} = Q_{lli}.
\end{array}}
\end{split}
\end{align}
where we have put 
\[
B^{\alpha \beta \gamma} = A^{\alpha \langle  \beta \gamma \rangle} = A^{\alpha \beta \gamma} - \, \frac{1}{4} \, A^{\alpha \mu \nu} g_{\mu \nu} g^{\beta \gamma}.
\]
The equations on the right side of  \eqref{Limiti} correspond to the classical general balance equation for $14$ fields reported in \cite{AMR2023}. 
In particular from \eqref{Limiti} and  \cite{AMR2023}, we have
\begin{align}\label{limitec}
\begin{split}
&  \lim_{c\rightarrow \infty, \, v_i \rightarrow 0}  \left( 2 c \, B^{k  0  i} -c^2 T^{ki} \right) = \hat{I}_{llki} = b_C \, \delta_{ki}, \\
& \lim_{c\rightarrow \infty, \, v_i \rightarrow 0}\left(\frac{4 }{c}\, B^{0  ll } -6   \, T^{0 0} + 3 c V^0\right) = \hat{H}_{ll }= 3 a_{C},\\
&  \lim_{c\rightarrow \infty, \, v_i \rightarrow 0}{\left(-4 I^{ij}\right)}\delta_{ij} =\hat{P}_{ll} =  3 a_{3C} \, \pi,\\
&  \lim_{c\rightarrow \infty, \, v_i \rightarrow 0}{\left(I^{<ij>_3}\right)}  =\hat{P}_{<ij>} = -  a_{2C}\,  t_{<ij>_3}, \\
&\lim_{c\rightarrow \infty, \, v_i \rightarrow 0}{\left(2 c I^{0j}\right)}  =\hat{Q}_{llj} =   a_{1C}\,  q_{j}
\end{split}
\end{align}
where $a_C$, $b_C$  and $a_{1C}, a_{2C}, a_{3C}$ denote the corresponding classical equilibrium function defined in \cite{AMR2023}.
Taking into account \eqref{Va} and \eqref{Tab} evaluated in equilibrium and \eqref{aaa}, \eqref{iii}, the \eqref{limitec} gives
\begin{align}\label{lim1}
\begin{split}
  \lim_{c\rightarrow \infty} a  = \frac{1}{4}\left(a_C + \rho c^2 +2 \rho \varepsilon\right), \qquad 
 \lim_{c\rightarrow \infty} b = p + \frac{b_C}{2 c^2},
\end{split}
\end{align}
and
\begin{equation}\label{lim2}
\lim_{c\rightarrow \infty} a_1 = \frac{ a_{1C}}{2 c^2}, \qquad 
\lim_{c\rightarrow \infty} a_2 = - a_{2C},\qquad
\lim_{c\rightarrow \infty} a_3 = \frac{ a_{3C}}{ c^2}.
\end{equation}
Inserting \eqref{lim1} and \eqref{lim2} into \eqref{ab0} and \eqref{bellino} we obtain that the classical functions $a_C, b_C, a_{1C}, a_{2C}, a_{3C}$ satisfy the same conditions of Theorem \ref{teorema1} of \cite{AMR2023}.

\section{Conclusions}
 Any hyperbolic theories of divergence form with $14$ fields for a generic gas non necessarily  rarefied can have a triple tensor in equilibrium  $A_E^{\alpha <\beta \gamma>}$ and a production term  $I^{<\beta \gamma>}$ compatible with Theorem 1 that is a necessary and sufficient condition for the first approximation of Maxwellian iteration to coincide with the Eckart system. 
   In  Theorem \ref{teorema2}  we proved the  property that any causal  relativistic theory compatible with an entropy principle automatically converges in the parabolic limit to the Eckart system. This means, roughly speaking, that the entropy principle remains valid in the limit when the relaxation time goes to zero, according to Maxwellian iteration. Therefore,  
the present study clarifies the RET theories in their parabolic limit and may be helpful to construct new causal   relativistic theories valid also in the range in which the kinetic theory cannot be applied such as  in dense gases. We have proved that in a theory not far from equilibrium, only two equilibrium functions are needed to construct such a theory.
\bigskip

\section*{Acknowledgments}
\small{ The author thanks   Heinrich Freist\"uhler for the stimulating discussion during his permanence at Konstanz University.    The work has  been carried out in the framework of the activities of the Italian National Group of Mathematical Physics of the Italian National Institute of High Mathematics GNFM/INdAM.}

\end{document}